\documentclass{article}

\usepackage{multicol,multirow,makecell,hhline}

\usepackage{booktabs,caption,amsmath}

\usepackage[flushleft]{threeparttable}

\usepackage{authblk,geometry,abstract,array}

\usepackage{graphicx,xcolor} 
\usepackage{multirow}

\title{Do we listen to what we are told? An empirical study on human behaviour during the COVID-19 pandemic: neural networks vs. regression analysis}

\author[1]{Yuxi Heluo\thanks{yuxi.heluo@univie.ac.at (corresponding author)}}

\author[2]{Kexin Wang\thanks{k.wang@hw.ac.uk}}

\author[3]{Charles W. Robson\thanks{charles.robson@tuni.fi}}

\affil[1]{Department of Business Decisions and Analytics, University of Vienna, Oskar-Morgenstern-Platz 1, 1090 Vienna, Austria}

\affil[2]{School of Engineering and Physical Sciences, Heriot-Watt University, Edinburgh EH14 4AS, UK}

\affil[3]{Physics Unit, Tampere University, Tampere FI-33720 Finland}

\date{}

\newcolumntype{C}[1]{>{\centering\arraybackslash}m{#1}}

\begin{document}

\maketitle

\begin{multicols}{2}
[
\begin{abstract}
In this work, we contribute the first visual open-source empirical study on human behaviour during the COVID-19 pandemic, in order to investigate how compliant a general population is to mask-wearing-related public-health policy. Object-detection-based convolutional neural networks, regression analysis and multilayer perceptrons are combined to analyse visual data of the Viennese public during 2020. We find that mask-wearing-related government regulations and public-transport announcements encouraged correct mask-wearing-behaviours during the COVID-19 pandemic. Importantly, changes in announcement and regulation contents led to heterogeneous effects on people's behaviour. Comparing the predictive power of regression analysis and neural networks, we demonstrate that the latter produces more accurate predictions of population reactions during the COVID-19 pandemic. Our use of regression modelling also allows us to unearth possible causal pathways underlying societal behaviour. Since our findings highlight the importance of appropriate communication contents, our results will facilitate more effective non-pharmaceutical interventions to be developed in future. Adding to the literature, we demonstrate that regression modelling and neural networks are not mutually exclusive but instead complement each other.
\end{abstract}
]

Pandemics have been a blight on humanity for centuries. Although the emergence and spread of coronavirus disease 2019 (COVID-19) has been deeply shocking, it was preceded by many other --- some far worse --- catastrophes: the ``Spanish flu'' of 1918-1919 killed more than the First World War, and the Black Death of the 14th century was responsible for the deaths of around half the population of Europe by some estimates \cite{Aberth}.

What is different about the most recent pandemic is the tools we now have. The modern, interconnected world is blessed with both advanced medical treatments as well as myriad forms of reliable and rapid communication. Not only do these advances allow pandemics to be attacked biologically --- by vaccinating the general population --- but they also enable safety information to be swiftly sent to the public, for example recommending optimal behaviours to reduce the spread of the disease.

Crucially, modern technology supports us in both communicating widely and rapidly about safe behaviour during a pandemic as well as \emph{quantitatively} studying the effectiveness of each intervention. By doing so, public-health communication may be greatly improved, with the potential to save many lives in future pandemics and other public health emergencies.

COVID-19 has been with us now for almost four years \cite{WHO_1}. During this time, policymakers around the world have implemented various public health strategies, including social distancing and mask wearing, to combat the threat. Empirical studies have provided evidence that the wearing of facial covers and maintaining a physical distance between people can indeed effectively reduce the risk of contagion \cite{WHO_2,Lancet,BMJ}.

The empirical literature has additionally shown that social distancing policies, such as lockdowns, reduced public mobility \cite{Heiler}, however, to the best of our knowledge, there is no clear evidence that mask-related policies encouraged correct mask-wearing behaviour in the general public. Combining object-detection-based convolutional neural networks (CNNs), regression analysis and multilayer perceptrons (MLPs), in this article we contribute the first results revealing to what extent a given populace responds compliantly to mask-wearing policies. Both government regulations and public transport announcements are considered.

Due to high population densities, public transport has been identified as a high-risk area for the spreading of COVID-19 \cite{Buja}. In this article, we investigate the association between public policy and human behaviour in Vienna, a major European city, during the year 2020. The public transport network of Vienna is extensive --- its usage accounts for approximately 38{\%} of total trips taken by individuals within the city \cite{Rasca}. This large daily flow of passengers makes Vienna an ideal case study for examining the role of transport-infrastructure-mediated-policy during the COVID-19 pandemic. The findings of this work should be generalizable to other major cities.

\begin{figure}
    \centering
    \includegraphics[scale=0.5]{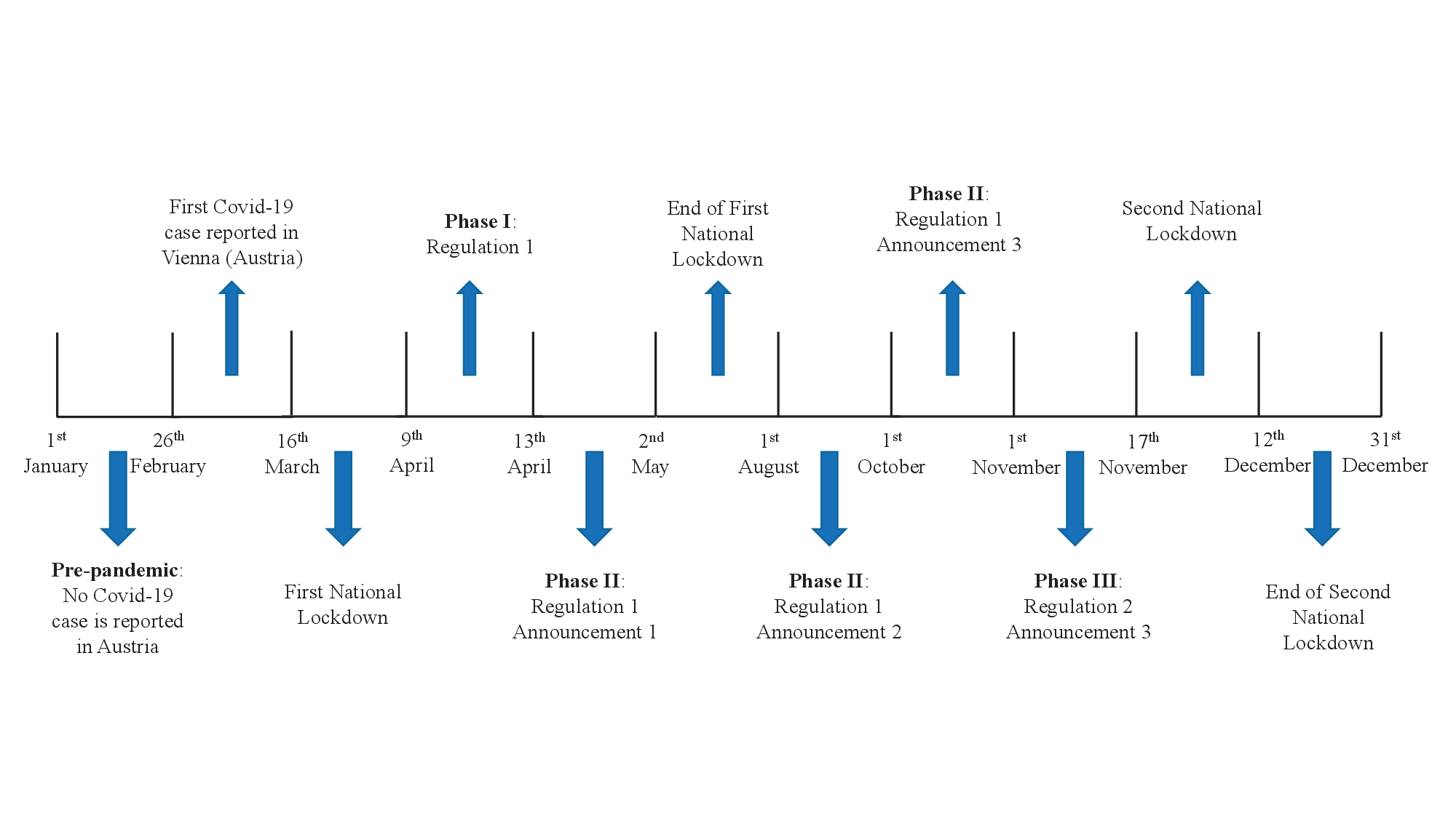}
    \caption{Timeline of COVID-19-preventive measurements in Vienna in 2020.}
    \label{fig:timeline}
\end{figure}

Beginning in April 2020, passengers on public transport and in stations in Vienna, Austria were required by the government to ``\emph{cover their mouth and nose with a mechanical protective}'' \cite{Fed}. In addition, between April 2020 and March 2023, the major Viennese public transport company, \emph{Wiener Linien}, introduced an audio announcement on transport and in stations and subsequently changed its content six times. Importantly, changes in the content of announcements did \emph{not} happen simultaneously with government regulation updates. In this article, we focus on the year 2020; see Figure \ref{fig:timeline} for a detailed timeline.

The fact that regulation and announcement changes did not overlap provides us with an invaluable opportunity to decouple the effects of government regulations and public transport announcements, thereby permitting a study of how each affected the mask-wearing behaviour of the Viennese public in 2020. We therefore propose the following research question: \emph{did mask-wearing-related government regulations and announcements on public transport and in stations affect people's mask-wearing behaviour during the COVID-19 pandemic in Vienna, and if so, how?}

In the past, studies of how policy affects societal behaviour have used proxy measures: in two different studies, mobile phone usage data was used to investigate mobility changes during COVID-19 \cite{Heiler,Levin}. These indirect studies of human behaviour are, of course, highly valuable, adding to our understanding of reactive behaviour during crises. However, it remains the case that an \emph{ideal} study of human behavioural responses during a pandemic would involve \emph{direct measures} of human actions. Not only do proxy studies suffer from intrinsic flaws but they are also severely limited in their scope: for example, it would be unfeasible to capture mask-wearing using people's mobile phone or other signal data.

In our approach, CNN technology is employed to process direct visual evidence of the mask-wearing behaviour of large groups of people. This allows us to avoid the pitfalls of proxy studies\footnote{For example, using mobile phone data to measure human mobility carries the potential risk of multi-counting individuals who possess more than one mobile device.} and to produce generalisable results. A subfield of artificial intelligence, machine learning is a powerful tool used to classify complex data and forecast future events \cite{IBM}. Despite the huge capabilities of machine learning models, many of them are ``black boxes'' \cite{Rudin,Mak}, i.e. the complicated internal processes generating outputs can be unclear to a user\footnote{For some recent work showing the dangers of this lack of interpretability, and how the situation can be improved, see Refs. \cite{DeGrave,Gardiner,Hu,Gouv}.}. Regression analysis, on the other hand, is a commonly used statistical method which models the relationships between output and input variables, with the advantage that the model's structure can be easily extracted.

In this work, we combine neural networks with regression analysis. By combining these two approaches, we can generate an accurate predictive model whilst simultaneously uncovering possible causal pathways underlying the system.

Two different neural networks are used in this work. The first (which we name ``Ada'') we manually train to recognize five different types of mask-wearing behaviour extracted from video data of the Viennese public. This behavioural information is then fed into the second of our networks (called ``John''), which is trained to predict people's mask-wearing behaviour based on certain inputs; these inputs we determine from our regression analysis. In this way the regression analysis complements the machine learning modelling.

As has been pointed out \cite{Mak}, many published studies make strong claims for the forecasting accuracy of their machine-learning models, without making a comparison with simple statistical methods. In our work, we quantitatively compare the prediction accuracy of our regression and neural network models, showing that the latter is a better forecaster of human behaviour in the context of our study.

In the following section, we briefly review the current state of research in the field and motivate and introduce our hypotheses. Then in the \emph{Data and methods} section, we explain our data collection process and our data analysis methodology. The subsequent \emph{Results} section gives the findings of our regression modelling and neural network analysis, as well as a comparison between the two approaches. Finally, in \emph{Discussion and conclusion}, we review our contributions, propose policy implications and suggest future work.

\section*{Research background and hypotheses}

The literature has shown that mask-wearing-related government policies were some of the most effective in reducing the spread of COVID-19 \cite{Haug}. Hence, these policies presumably increased the proportion of society wearing the required masks in the correct fashion, covering both their nose and mouth, leading in turn to reduced transmission of the SARS-CoV-2 virus. We therefore propose the following hypothesis:
\begin{center}
\emph{Hypothesis 1 (H1): Regulations 1 and 2 increased the percentage of people wearing masks correctly.}
\end{center}
See Table \ref{table:regs_anns} for the contents of Regulations 1 and 2, enforced by the Viennese government in 2020.

It is known that adequate communication tools combined with appropriate message content can affect people's \emph{intended} or \emph{self-reported} behaviour \cite{Lunn,Sasaki,Falco,Barari,Everett}. Building on this, two of the present authors have empirically shown that messaging does in fact change people's behaviour: specifically, it was demonstrated that underground announcements influenced public transport mobility in Vienna during the pandemic and that these effects were heterogeneous across different announcement contents \cite{Our_first_paper}. Consequently, we expect that underground mask-related announcements encouraged people to wear masks in the required manner. We further predict that different underground announcements affected people's mask-wearing behaviours in different ways --- this would be in line with the communications literature \cite{Lunn,Sasaki}, which has shown that people's reactive behaviour depends on the wording and elements of the communicated message. We thus hypothesize the following:
\begin{center}
\emph{Hypothesis 2a (H2a): Underground announcements increased the percentage of people wearing masks correctly.}
\end{center}
\begin{center}
\emph{Hypothesis 2b (H2b): The effects of different announcements on people’s mask-wearing behaviours were heterogeneous.}
\end{center}
See Table \ref{table:regs_anns} for the contents of these announcements.

\section*{Data and methods}

\paragraph{Data---}

We built a time-series dataset which includes daily street-interview data filmed in Vienna, collected from Austrian news agencies in 2020. By analysing these open-source videos using an object-detection-based machine-learning network, we identified five different mask-wearing behaviours: 
\begin{enumerate}
\item Not wearing any facial covers.
\item Wearing a disposable medical mask or respirator mask, but not covering both nose and mouth.
\item Wearing a disposable medical mask (e.g. a surgical mask) and covering both nose and mouth.
\item Wearing a respirator mask (e.g. FFP2 or N95) and covering both nose and mouth.
\item Wearing a facial cover, but not a disposable medical mask or respirator mask.
\end{enumerate}
To facilitate easy reading, we label the above behaviours as \emph{no facial cover}, \emph{wearing a mask wrongly}, \emph{disposable mask}, \emph{respirator mask}, and \emph{other facial cover}, respectively. A disposable medical mask is also known as a surgical mask and is defined as ``a loose-fitting, disposable device that creates a physical barrier between the mouth and nose of the wearer and potential contaminants in the immediate environment''; a respirator mask is ``a respiratory protective device designed to achieve a very close facial fit and very efficient filtration of airborne particles'' \cite{FDA}.

Here, we employ one of the most popular vision artificial intelligences, You Only Look Once version 5 (YOLOv5) \cite{Yolov5, Yolov5Paper, Yolov5App1, Yolov5App2, Yolov5App3, Yolov5Mask1, Yolov5Mask2, Yolov5Mask3}, to identify the percentage of different mask-wearing behaviours in the collected media sources, where LabelImg \cite{LabelImg, LabelImgApp1, LabelImgApp2}, a data annotation tool, is used to label the area of interest in each selected video frame, therefore improving the decision accuracy of YOLOv5.

The Viennese government policy data was collected from the Austrian Federal Law Gazette. The interviewing locations and announcement information were identified and collected manually by our group members.


\paragraph{Mask Detection (Ada)---}

The Yolov5 artificial intelligence is employed by Ada for mask detection. This cutting-edge object-detection neural network provides high accuracy, rapid inference speed, and user-friendliness. It has been widely applied in various computer vision tasks, notably in real-time object detection \cite{Yolov5Paper} and mask detection \cite{Yolov5Mask1,Yolov5Mask2,Yolov5Mask3}. Meanwhile, LabelImg, a proven effective tool for object annotation in real-time applications \cite{LabelImgYolo1, LabelImgYolo2}, is used for annotating objects during the training, testing, and detection phases of Yolov5. The default settings of Yolov5 were adopted, specifically using the `yolov5s' model as the pre-trained checkpoint, which is optimized for small and medium-sized objects \cite{Yolov5}.

Prior to Ada’s deployment in detecting mask wearing behaviour in our street-interview data, she underwent training with 960 open-source images containing mask-wearing information and was tested using 373 open-source images. The results of this testing phase are detailed in Table \ref{table:Ada_data}.

Our video dataset contains footage spanning 293 days during the year 2020. Yolov5 and LabelImg detected mask-wearing behaviours in 284 days. On average, approximately 104 behaviours were identified per day.


\paragraph{Regression analysis---}

Our dependent variables are the percentage of people's engagement in each mask-wearing behaviour in each collected video. On average, in each day, 58.65\% of people did not wear any facial covers, and 27.88\% of people wore a required mask and covered both nose and mouth. Independent variables include mask-related government regulations, which we label \emph{Regulations 1} and \emph{2}, and underground announcements, denoted \emph{Announcements 1}, \emph{2} and \emph{3}. All independent variables are dummy variables, meaning if a mask-wearing-related regulation or announcement is present, our independent variables are equal to one, otherwise, they are equal to zero. 

\begin{figure}
    \centering
    \includegraphics[scale=0.85]{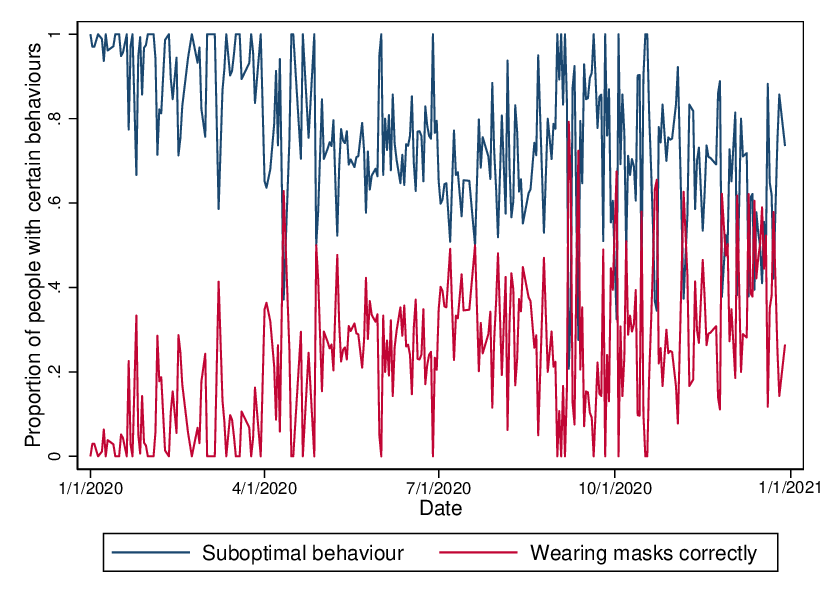}
    \caption{People's mask wearing behaviour against time (non-stationary). Note that Suboptimal behaviour refers to the sum of not wearing any facial covers, wearing masks wrongly, and wearing other facial covers. Wearing masks correctly refers to the sum of wearing disposable medical masks and respirator masks (covering both nose and mouth).}
    \label{fig:nonstat}
\end{figure}

\begin{figure}
    \centering
    \includegraphics[scale=0.85]{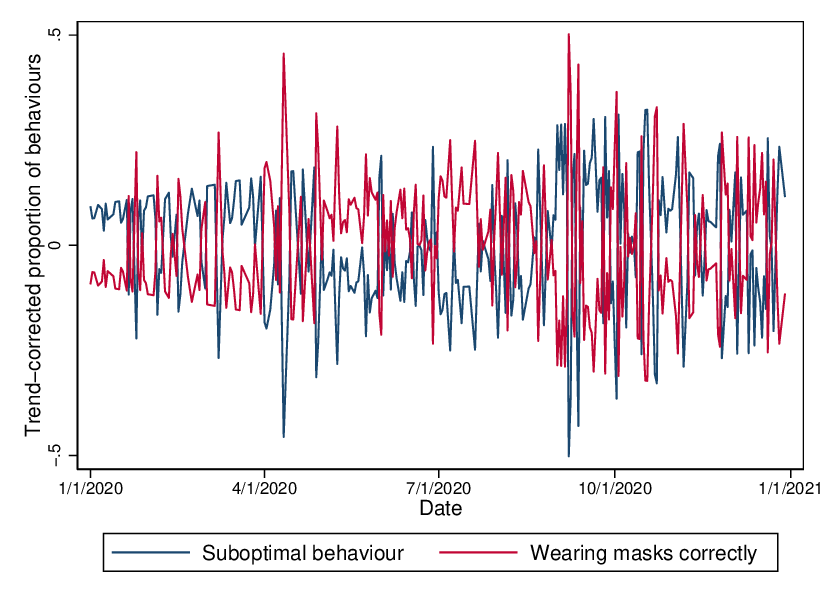}
    \caption{People's mask wearing behaviour against time (stationary). Note that Suboptimal behaviour refers to the sum of not wearing any facial covers, wearing masks wrongly, and wearing other facial covers.Wearing masks correctly refers to the sum of wearing disposable medical masks and respirator masks (covering both nose and mouth).}
    \label{fig:station}
\end{figure}

We plotted the proportion of a given mask-wearing behaviour against time in Figure \ref{fig:nonstat}. All dates in this paper are in the format mm/dd/yyyy. Since the data observations are not independent of time, we adopted model fitting techniques to achieve stationarity \cite{Wooldridge} (see Figure \ref{fig:station}).

Next, we incorporated the \emph{distance} between filming locations and the nearest underground station, the \emph{minimum and maximum temperatures} of the day, as well as the geographical \emph{districts} where the interview videos were recorded, as control variables. Given the government regulations mandating the use of masks in public transport stations, we expect that a decrease in the distance between people's locations and the nearest underground station will result in a higher proportion of individuals wearing masks. 

Empirical research has demonstrated that in hotter weather, individuals feel uncomfortable wearing masks \cite{Milošević}. Hence, we expect that increases in temperature is negatively associated with the percentage of people wearing masks.

\begin{figure}
    \centering
    \includegraphics[scale=0.5]{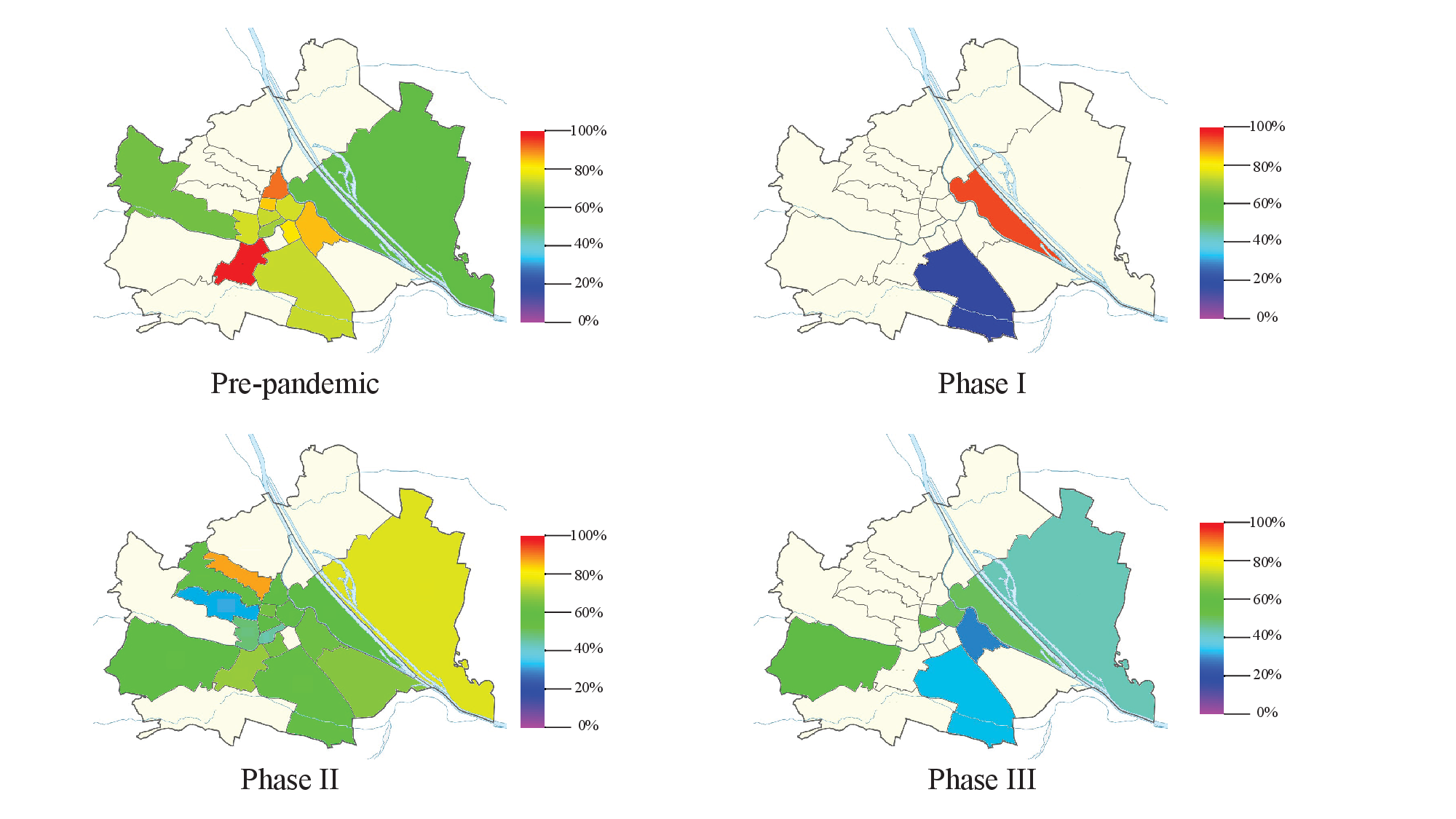}
    \caption{{Percentage of people not wearing any facial covers across different districts in 2020. Note that districts with no colour denote no data.}}
    \label{fig:Heatmap_no_cover_all}
\end{figure}

Mapping people's mask-wearing behaviours across all geographical districts in Vienna, we demonstrate that these behaviours exhibit heterogeneity across districts (see an example in Figure \ref{fig:Heatmap_no_cover_all}). Thus, we generate dummy variables for different districts to consider the effect of location on mask wearing behaviours. Due to the small number of observation days (17) in Phase I, we exclude district dummies in Phase I, to prevent overfitting. Table \ref{table:desc_stats} shows descriptive statistics.

\paragraph{Machine learning analysis (John)---}

The multilayer perceptron (MLP) is a type of neural network characterized by straightforward connections between each layer. Each layer links to the next via weighted links, and data moves from input to output layers through hidden layers without looping back. Neurons in these hidden layers perform mathematical functions. The activation function in these nodes provides nonlinear fitting capability. MLPs were widely used in the communication field as it excels in adapting to the non-linear and time-varying factors within its learning and prediction object \cite{CommMLP1, CommMLP2, CommMLP3}. Due to the presence of nonlinear factors within the behaviour analysis task, including the temperature and distance, John equips with a MLP to perform training and prediction.

The MLP is structured with an input layer, a hidden layer, and an output layer, forming a feedforward neural network. This output layer comprises five neurons, each representing a distinct mask-wearing behaviour. Given that the input for the MLP is a time-series dataset with a size of fewer than 366 groups (attributable to invalid data collection), it is important to adopt a minimalist MLP configuration. This approach aims to mitigate overfitting within the network while preserving the MLP's accuracy in learning and prediction. Additionally, employing a single hidden layer simplifies the process of delineating the relevance of different inputs with respect to each output. The hidden layer consists of 50 neurons to ensure robust predictive capabilities.

In line with our regression model, the input variables for our neural network are \emph{distance}, \emph{districts}, \emph{minimum and maximum temperatures}, \emph{Announcements 1}, \emph{2} and \emph{3}, and \emph{Regulations 1} and \emph{2}. The input layer also incorporates temporal memory elements. For instance, an MLP configured with a three-day temporal memory integrates the three preceding prediction outcomes alongside the input variables. This inclusion enhances the model's capacity to factor in recent trends and developments in its analysis.

\begin{figure}
    \centering
    \includegraphics[scale=0.5]{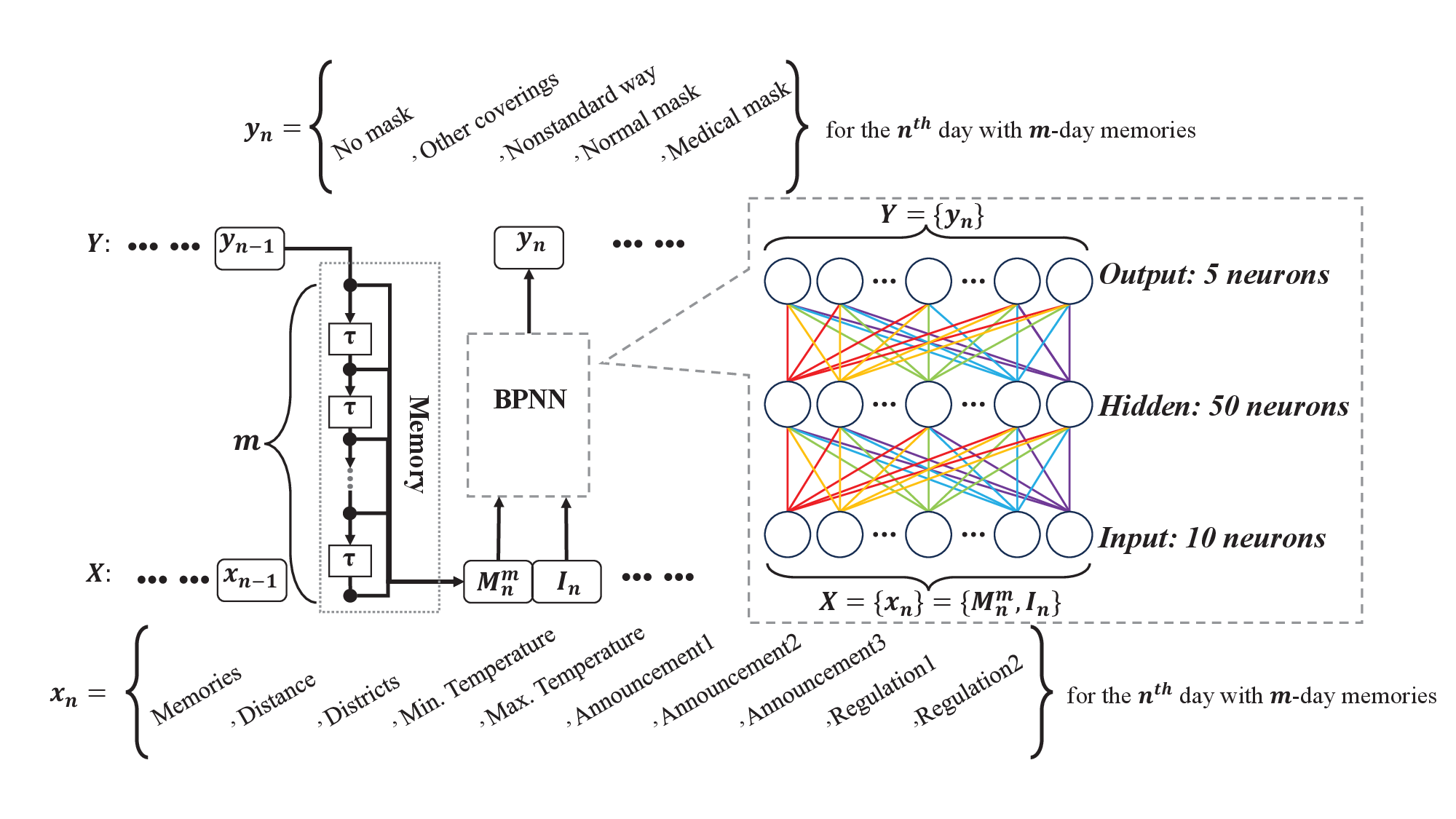}
    \caption{{Structure of MLP (John).}}
    \label{fig:John.}
\end{figure}

Sprengholz et al. \cite{Sprengholz} demonstrated an association between individuals' mask-wearing behaviours in 2022 and their behaviours in 2020 and 2021. Using mobility as a proxy of people's behaviour, Heiler et al. \cite{Heiler} observed a weekly trend of public transport usage during the COVID-19 pandemic. To consider these associations, we incorporated memory windows in the MLP. However, since there is no clear guidance regarding the length of the memory for mask wearing behaviours, we choose 0-, 3-, 7-, and 11-day memory windows to capture possible trends in people's mask wearing behaviours (see Figure \ref{fig:John.} for the structure of MLP). 

To assess the predictive power of our MLP and compare it with our regression models, we employ four common indices: 1) mean squared error (MSE),
\begin{equation*}
    \mathrm{MSE} = \frac{1}{m}\sum_{i=1}^{m}\left( X_{i} - Y_{i} \right)^{2} ,
\end{equation*}
2) root-mean-squared deviation (RMSE), 
\begin{equation*}
    \mathrm{RMSE} = \sqrt{\frac{1}{m}\sum_{i=1}^{m}\left( X_{i} - Y_{i} \right)^{2}} ,
\end{equation*}
3) mean absolute error (MAE) , 
\begin{equation*}
    \mathrm{MAE} = \frac{1}{m}\sum_{i=1}^{m} \left| X_i - Y_i \right| ,
\end{equation*}
and 4) the symmetric mean absolute percentage error (sMAPE),
\begin{equation*}
    \mathrm{sMAPE} = \frac{2}{m} \sum_{i=1}^{m} \frac{\left| X_i - Y_i \right|}{\left| X_i \right| + \left| Y_i \right|} ,
\end{equation*}
where in all equations $X_i$ is the predicted $i$th value, $Y_i$ is the actual $i$th value
\cite{Hamzaçebi,Kelley,Sammut_a,Sammut_b,Hyndman,Chen}.

\section*{Results}

\paragraph{Regression analysis---}

We split our regression analyses into three phases based on the timing of each regulation and announcement. In Phase I, we detected 1,305 people across 17 days. Table \ref{table:OLS_I} shows how \emph{Regulation 1} influenced people's mask-wearing behaviour in Phase I. We found that \emph{Regulation 1} encourages people to wear masks as required. In detail, the presence of \emph{Regulation 1} increased the percentage of people wearing a respirator mask by 0.13{\%}; this result is statistically significant at the 5{\%} level. The average number of passengers using the public transport system to commute in Vienna was 1.57 million per day in 2020 \cite{WL}. In other words, on average, \emph{Regulation 1} increased the number of people wearing respirator masks correctly by 2,041 per day. 

In Phase II, we detected 22,636 people across 140 days. We found that the presence of any underground announcement decreased the percentage of people not wearing any facial covers by 1.641{\%} (i.e. by 25,763 people on an average day) --- this result is statistically significant at the 1{\%} level. The presence of an announcement also increased the percentage of people who wore disposable medical masks and covered both nose and mouth by 0.353{\%} (that is, by around 5,542 people per day), and it increased the percentage of people who wore a respirator mask and covered both nose and mouth by 0.811{\%} (by around 12,733 people per day). These results are statistically significant at the 5{\%} and 10{\%} level, respectively; see Table \ref{table:OLS_II}.

In the marginal effect model, we found that such positive effects on mask-wearing behaviours were sustained even after individuals were up to 220 meters away from an underground station (see Table \ref{table:margeff_II}). Interestingly, contrasted with \emph{Announcement 3}, we also found that \emph{Announcement 1} did not affect people's mask wearing behaviour; see Table \ref{table:margeff_123}. Compared with \emph{Announcement 3}, \emph{Announcement 2} decreased the percentage of people wearing a respirator mask and covering both nose and mouth by between 0.036{\%} and 0.038{\%} (around 565 to 597 people). Finally, compared with \emph{Announcement 2}, \emph{Announcement 3} increased the percentage of people wearing a disposable medical mask and a respirator mask and covering both nose and mouth by 0.154{\%} and 0.032{\%}, respectively (which is around 2,418 and 502 people). It also decreased percentage of people not wearing any facial covers by 0.128{\%} (by around 2010 people); see Table \ref{table:margeff_123}.

In Phase III, we detected 3,754 people across 34 days and found that, compared with \emph{Regulation 1}, \emph{Regulation 2} decreased the percentage of people wearing a disposable medical mask by 0.132{\%} (by around 2,072 people) --- this result is statistically significant at the 10{\%} level; see Table \ref{table:OLS_III}.

\paragraph{Machine learning analysis---}

In this section, we introduce the training and testing of Ada and John. The testing, conducted with 373 open-source images, yielded results as listed in Table \ref{table:Ada_data}. The detection accuracy for the five distinct mask-wearing behaviours was 95{\%}, 100{\%}, 99{\%}, 98{\%}, and 100{\%}. Incorrect detections in the `no facial covers' case could be attributed to lighting variation in the training and testing photographs. Additionally, the confusion between disposable medical masks and respirator masks may be linked to posture or gestures in the dataset.

\begin{figure}
    \centering
    \includegraphics[scale=0.5]{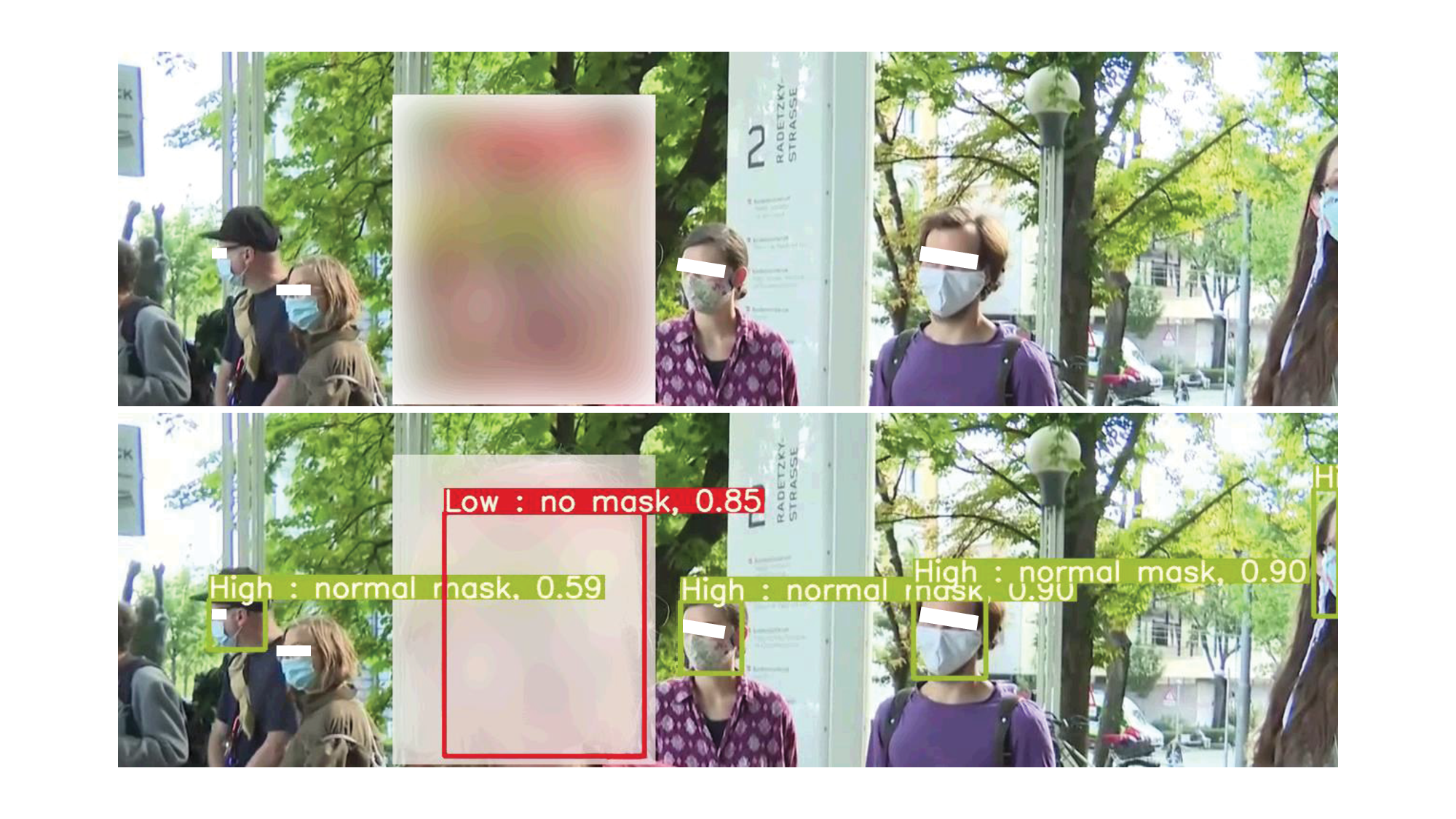}
    \caption{An example of the mask-detection output. To protect the interviewee's identity, we have blurred their face.}
    \label{fig:MaskDetection}
\end{figure}

Figure \ref{fig:MaskDetection} shows an example of Ada's detection output. The video, filmed by an Austrian news agency, is of an interview with several bystanders in the background. Ada successfully identified the mask-wearing status of these individuals. It is important to note that interviewees are often required to remove their masks during interviews. Hence, the interviewee was excluded in the detection results.

To train John, we filtered out low-quality videos. In total, we have 225 training and testing groups. After 1,000 iterations, our MLP has reached its floor. Table \ref{table:MLP_vs_Reg} shows the prediction accuracy of John.

\paragraph{Comparison between regression and MLP results---}

In this section, we compare the predictive power between regression and MLP analyses. In our regression models, there is no auto-correlation ($P_{\textrm{Durbin's \ alternative \ tests}} >$ 0.1). Hence, we take our MLP results with no memory window to compare with regression's predictions. We estimate the MSE, RMSE, MAE, sMAPE indices. While Chicco et al. \cite{Chicco} argue that, compared with other indices, R-squared is a more informative and truthful metric for evaluating regression analyses, Spiess and Neumeyer \cite{Spiess} demonstrated that R-squared is not applicable in a non-linear regime. Given that our MLP model operates in a non-linear fashion, it is inappropriate to use R-squared to assess its performance.

We found that in Phases II and III, these four indices for our MLP are smaller than those for our regression, indicating that our MLP gives better predictions than regression in these two phases (see Table \ref{table:MLP_vs_Reg}). Note that the numbers in bold in Table \ref{table:MLP_vs_Reg}) stand for higher prediction accuracy.

For predictions of the percentage of people wearing disposable masks in Phase I, regression exhibits the same MSE and RMSE ($MSE_{\textrm{regression}}$=0.000, and $RMSE_{\textrm{regression}}$=0.000) as MLP ($MSE_{\textrm{MLP}}$=0.000) and $RMSE_{\textrm{MLP}}$= 0.000). However, MSE and RMSE are sensitive to the scale of observations and outliers, as is the case in our Phase I sample; therefore, they are less valid indicators in this context \cite{Chicco}. Further, our multilayer perceptron model has smaller MAE and sMAPE, indicating a higher prediction accuracy than regression in Phase I.

In summary, our MLP model gives better predictions than our regression model.

\paragraph{Robustness tests---}

It could be argued that the heterogeneous effects of the three different underground announcements on people’s mask-wearing behaviours might not be statistically significant. We therefore run t-tests to validate the statistical significance (see Table \ref{table:ttest123}). We found that the effects of \emph{Announcements 2} and \emph{3} are statistically significantly different from each other, but not those between \emph{Announcements 1} and \emph{2}. 

Considering that people may not respond to an announcement on the first day of its realisation, we include a one-day lagged term in our regression models in Phase II to take this into account. We find that our results are consistent.

We conduct three sensitivity analyses for our regression models. First, to generalise people's mask-wearing behaviours, we ranked these behaviours based on their effectiveness at preventing COVID-19 spreading (see Table \ref{table:rank}). We then defined a new term, \emph{Strictness Index}, using the following equation:
\begin{equation*}
\begin{aligned}
& \emph{Strictness Index} = \\
& \sum_{i=1}^{5} \Bigg[ \Bigg( \frac{\text{Number of people with behavior } i}{\text{Total number of people detected}} \Bigg) \\
& \times \text{Rank of behavior } i \Bigg] .
\end{aligned}
\end{equation*}

This index measures how strict people are regarding mask-wearing during COVID-19.  We found that, compared with \emph{Regulation 1}, \emph{Regulation 2} increased people’s \emph{Strictness Index} (see Table \ref{table:strictness}). In Phase II, in the marginal effect model, we found that, compared with  \emph{Announcements 1} or \emph{2}, \emph{Announcement 3} increased people’s \emph{Strictness Index} (see Table \ref{table:marg_strict}). Therefore, our findings are robust.

For our second sensitivity analysis, instead of district dummies, we use district-related characteristics to capture locational differences in Vienna. We use data on \emph{annual income}, \emph{percentage of people having a bachelor's degree}, \emph{number of people per hectare}, and \emph{percentage of non-immigrants} to capture district differences. Table \ref{table:margin_district} shows that, compared with \emph{Announcements 1} or \emph{2}, \emph{Announcement 3} encouraged mask-wearing behaviour. In contrast with \emph{Regulation 1}, \emph{Regulation 2} led to a lower percentage of people who wore disposable masks correctly in Phase III (see also Table \ref{table:Robust_district}). Therefore, our findings are robust.

Thirdly, we acknowledge that individuals may interpret the same temperature differently across different seasons. For instance, a temperature of 15 degrees Celsius may be perceived as cool during the summer in Vienna (where summer temperatures range between 9 and 34 degrees Celsius), whilst the same temperature may be considered warm during the winter (with winter temperatures ranging between -7.2 and 18 degrees Celsius). To address this concern, we included seasonal dummies in our regression models: \emph{Spring}, \emph{Summer}, \emph{Autumn}, and \emph{Winter}. Consistent with our findings, Table \ref{table:Robust_season_overall} shows that \emph{Regulation 1} or the presence of any announcement encourages more people to wear masks as required, whereas, contrasted with \emph{Regulation 1}, \emph{Regulation 2} decreased the percentage of people who wore disposable masks correctly. Therefore, our findings are robust.

We ran t-tests on the MLP-predicted results and real-life data, finding that there are no statistically significant differences between them. Hence, our MLP predictions are accurate. 

We also compare the predictive accuracy of our MLP models with different memory windows using MSE, RMSE, MAE, and sMAPE. We found that MLP with a memory window of zero days provides the most accurate prediction in each phase and overall (see an example comparison in Table \ref{table:NN_Memory window}). In other words, the proportion of people who wore masks at time $t$ in our dataset might not be correlated with the percentage of people who wore masks in time $t-n$ ($n$ stands for a time lag).

In the field of economics and behavioural studies, compared with MLP, random forest (RF) is more popular \cite{Cheng,Tanaka}. 
In our study, we conducted a comparative analysis of the training and prediction performance between MLP and RF. For the RF model, we manually set the number of trees to nine inputs, corresponding to the number of input variables. The criteria for splitting and the number of internal nodes were automatically adjusted during the training process. The comparative results, presented in  Table \ref{table:NN_VS_RF} and Table \ref{table:RF_vs_MLP_time_node}, include metrics such as time consumption for training and prediction, the number of terminal nodes utilized in prediction, and prediction accuracy. The evaluation was performed on hardware equipped with “Intel(R) Core(TM) i5-8250U” processor and 16.0 gigabytes of RAM.

Our findings indicate that, in the context of mask-wearing behavior analysis, the RF model requires more time for training and utilizes a greater number of terminal nodes than the MLP. At the same time, it yields less accurate predictions. This result can be attributed to the use of the hyperbolic tangent function in MLP as its activation function, which is absent in the RF model. This nonlinear activation function endows the network with the capability to effectively capture the nonlinearity between inputs and outputs. To accommodate this nonlinearity, the RF model necessitates a more intricate structure with additional leaves. However, such a complex structure demands extensive computational resources and a larger dataset to achieve adequate training and prediction, which is impractical in this application.


\section*{Discussion and conclusion}

Modern technologies give humanity a powerful tool with which to manage public health catastrophes. In this article, we combine neural network and regression modelling to study the reactive behaviour of a given population, specifically Vienna in the year 2020. This allows us to examine the efficacy of various government regulations and public transport announcements

Street interviews throughout the year 2020 were collected and analysed by an object-detection-based convolutional neural network, permitting us to empirically assess the mask-wearing behaviours of a large number of individuals in a natural setting.

We find that the presence of a mask-wearing-related government regulation encourages a populace to wear a disposable medical or respirator mask, covering both their mouth and nose. We also discover that compared with the time period in which no announcements are present, the time period where there is a public transport announcement led to a higher proportion of people wearing disposable or respirator masks as required and, further, reduced the proportion of people not wearing any facial covers. It is also found that the effects of announcements on people's behaviour are heterogeneous, specifically \emph{Announcement 3} is more effective than \emph{Announcements 1} and \emph{2}.

Based on our regression modelling, we chose a set of input variables for our multilayer perceptron, resulting in highly accurate predictions of people's behaviours. Our work contributes, as far we are aware, the first study using regression modelling to optimally choose a series of input variables for neural network modelling. Due to our highly accurate MLP prediction results, we speculate that this approach to choosing input variables is constructive. We also believe that our study contains the first comparison between the predictive powers of linear regression and neural networks.

Adding to the literature of behavioural economics, we contribute here a new, direct measurement of human behaviour during the COVID-19 pandemic. Compared with previous proxy studies of a populace --- for example via mobile phone signal data or self-reporting --- our study captures and analyses visual evidence of a population's response to public policy. We also point out that open source videos are a rich stream of information in the study of reactive human behaviour.

Compared with past studies, which were mainly conducted in the laboratory, a natural experiment setting is studied here to assess policy effectiveness.

We highlight here the feasibility of using MLPs in the behavioural sciences. Due to the fact that relationships between ``input'' and ``output'' variables in real-world systems are highly nonlinear, using multilayer perceptrons (with nonlinear activation functions) is preferable to the commonly-used random forest modelling. We contribute a brief comparison of the structure, the required computational resources and predictive power of random forest and MLP modelling, showing that MLP provides more accurate predictions in a shorter time, using fewer computational resources.



Based on our results, we propose the following suggestions to policymakers: firstly, we recommend that, during a public health emergency, issuing behavioural regulations should be made a high priority; secondly, since we find that public transport announcements can effect people's behaviour, we suggest that government's take advantage of the existing infrastructure to disseminate critical information during crises; thirdly, as the effectiveness of policy depends crucially on its structure (e.g. wording and elements), we stress that messages should be tailored in an appropriate manner.

For economists, our work shows that regression and neural network modelling can work in tandem, and are not mutually exclusive. Traditional methods such as regression can be used to reveal possible causal pathways underlying a system, which then aids in the construction of machine-learning models, the latter improving system prediction accuracy.

We provide evidence here that neural networks are powerful and effective tools in the modelling and prediction of human behaviour, even when working with limited datasets and resources. It is our hope that researchers in the machine-learning field will continue to study public health and policy using neural networks.

We acknowledge that the spread of COVID-19 can affect people's mask-wearing behaviours, and vice versa. However, this is not our focus in this article, and we will consider this feedback loop in our future work. Our current study only considers the presence of public policy which, within our current methodology, is difficult to generalize. In the future, we will adopt text-mining techniques to rank the strictness of each policy to address this issue.

Finally, our current work consists of Viennese data from just the year 2020. For the next step, we will collect and analyze more video data from Vienna, and from other countries, also from 2021 and 2022. These larger datasets will aid us in generalizing our results and improving our model performance in the future.

\section*{Acknowledgments}

The authors would like to thank Professor Oliver Fabel, Dr. Xu Wang, Professor Rudolf Vetschera, Dr. Bolin Mao, Dr. Pruschak Gernot and Dr. Ayseg{\" u}l Engin for valuable discussions. We also would like to thank Zhaokang Zhou, Xiaotian Gao, Jiashu Yang, Kilian Kie{\ss}ling, Christina Molnar, and Anna Smekal for data processing.


\section*{Competing interests}

The authors declare no competing interests.

\end{multicols}

\newpage

\bgroup
\def\arraystretch{1.5}
\begin{table}[ht]
\centering
\caption{Government regulations and public transport announcements in Vienna in 2020}
\begin{tabular}{ |p{3cm}||p{3cm}|p{3cm}|p{3cm}|  }
 \hline
 \multicolumn{4}{|c|}{Vienna government regulations regarding mask wearing} \\
 \hline \hline
  \multicolumn{2}{|c|}{Month regulation imposed} & \multicolumn{2}{|c|}{Content of regulation} \\
 \hline
  \multicolumn{2}{|c|}{April 2020} & \multicolumn{2}{|c|}{\makecell{\emph{Passengers on public transport and in stations have to} \\ \emph{cover their mouths and noses with mechanical protection.}}} \\
 \hline
\multicolumn{2}{|c|}{November 2020} & \multicolumn{2}{|c|}{\makecell{\emph{Passengers have to wear a close-fitting mechanical} \\ \emph{protective device to cover their mouths and noses.}}} \\
\hline \hline
 \multicolumn{4}{|c|}{Wiener Linien announcements} \\
 \hline \hline
  \multicolumn{2}{|c|}{Month announcement started} & \multicolumn{2}{|c|}{Content of announcement} \\
 \hline
  \multicolumn{2}{|c|}{April 2020} & \multicolumn{2}{|c|}{\makecell{\emph{Dear passenger, please cover your nose and mouth in} \\ \emph{underground stations and when you are in public transport.}}} \\
 \hline
\multicolumn{2}{|c|}{August 2020} & \multicolumn{2}{|c|}{\makecell{\emph{Dear passenger, please cover your nose and mouth in} \\ \emph{underground stations and when you are in public transport.} \\ \emph{You know what? Your nose needs protection too.} }}  \\
\hline
\multicolumn{2}{|c|}{October 2020} & \multicolumn{2}{|c|}{\makecell{\emph{Dear passengers, cover your nose and mouth.}}} \\
\hline
\end{tabular}
\label{table:regs_anns}
\end{table}
\egroup

\newpage

\bgroup
\def\arraystretch{1.5}
\begin{table}[ht]
\centering
\caption{Object-detection-based convolutional neural network (Ada) training result. Note that Case 1 stands for no facial covers, Case 2 is wearing mask wrongly, Case 3 is wearing disposable masks, Case 4 is wearing respirator masks, and Case 5 is wearing other facial covers.}
\begin{tabular}{ |C{4cm}|C{2.4cm}|C{1.6cm}|C{1.6cm}|C{1.6cm}|C{1.6cm}|C{1.6cm}|}
 \hline
 \multicolumn{2}{|c|}{\multirow{2}{*}{Testing accuracy}}   & \multicolumn{5}{c|}{Actual}                                                                                                   \\ \cline{3-7} 
\multicolumn{2}{|c|}{}                                    & \multicolumn{1}{c|}{Case 1} & \multicolumn{1}{c|}{Case 2} & \multicolumn{1}{c|}{Case 3} & \multicolumn{1}{c|}{Case 4} & Case 5 \\ \hline
\multicolumn{1}{|c|}{\multirow{5}{*}{Predicted}} & Case 1 & \multicolumn{1}{c|}{0.95}  & \multicolumn{1}{c|}{}       & \multicolumn{1}{c|}{}       & \multicolumn{1}{c|}{}       &        \\ \cline{2-7} 
\multicolumn{1}{|c|}{}                           & Case 2 & \multicolumn{1}{c|}{0.03}  & \multicolumn{1}{c|}{1.00}   & \multicolumn{1}{c|}{}       & \multicolumn{1}{c|}{}       &        \\ \cline{2-7} 
\multicolumn{1}{|c|}{}                           & Case 3 & \multicolumn{1}{c|}{0.01}  & \multicolumn{1}{c|}{}       & \multicolumn{1}{c|}{0.99}   & \multicolumn{1}{c|}{0.02}   &        \\ \cline{2-7} 
\multicolumn{1}{|c|}{}                           & Case 4 & \multicolumn{1}{c|}{0.00}  & \multicolumn{1}{c|}{}       & \multicolumn{1}{c|}{0.01}   & \multicolumn{1}{c|}{0.98}   &        \\ \cline{2-7} 
\multicolumn{1}{|c|}{}                           & Case 5 & \multicolumn{1}{c|}{0.01}  & \multicolumn{1}{c|}{}       & \multicolumn{1}{c|}{}       & \multicolumn{1}{c|}{}       & 1.00   \\ \hline
\end{tabular}
\label{table:Ada_data}
\end{table}
\egroup

\newpage

\bgroup
\def\arraystretch{1.5}
\begin{table}[ht]
\centering
\caption{Descriptive statistics}
\begin{tabular}{ |C{4cm}|C{2.4cm}|C{1.6cm}|C{2cm}|C{1.6cm}|C{1.6cm}|  }
 \hline
 Variables & Observations & Mean & Std. Dev. & Min. & Max. \\
 \hline \hline
 Number of people not wearing any facial covers & 17,908 & 61.119 & 63.809 & 0 & 395 \\
 \hline
 Number of people wearing mask wrongly & 1,818 & 6.205 & 9.299 & 0 & 49 \\
 \hline
 Number of people wearing disposable mask & 6,878 & 23.474 & 27.0193 & 0 & 152 \\
 \hline
 Number of people wearing respirator mask & 1,842 & 6.287 & 10.486 & 0 & 76 \\
 \hline
 Number of people wearing other facial covers & 2,141 & 7.307 & 9.299 & 0 & 48 \\
 \hline
 Number of people detected & 30,587 & 104.393 & 104.701 & 0 & 503 \\
 \hline
 Announcement 1 & 366 & 0.301 & 0.459 & 0 & 1 \\
 \hline
 Announcement 2 & 366 & 0.167 & 0.373 & 0 & 1 \\
 \hline
 Announcement 3 & 366 & 0.251 & 0.434 & 0 & 1 \\
 \hline
 Regulation 1 & 366 & 0.563 & 0.497 & 0 & 1 \\
 \hline
 Regulation 2 & 366 & 0.167 & 0.373 & 0 & 1 \\
 \hline
 Min. temperature & 366 & 6.451 & 6.930 & -8 & 20 \\
 \hline
 Max. temperature & 366 & 16.073 & 8.887 & -1 & 34 \\
 \hline
 Distance & 238 & 408.882 & 440.689 & 0 & 2,800 \\
 \hline
\end{tabular}
\label{table:desc_stats}
\end{table}
\egroup

\newpage

\bgroup
\def\arraystretch{1.5}
\begin{table}[ht]
\begin{threeparttable}
\centering
\caption{OLS regressions for Phase I}
\begin{tabular}{ C{3.5cm}C{2.1cm}C{2.1cm}C{2.1cm}C{2.1cm}C{2.1cm}  }
 \hline 
  & No facial covers & Wearing masks wrongly & Disposable masks & Respirator masks & Other facial covers \\
 \hline
  & (1) & (2) & (3) & (4) & (5) \\
 \hline
 Regulation 1 & -0.012 & -0.076 & 0.053 & 0.130$^{**}$ & -0.095 \\
 
  & (0.956) & (0.190) & (0.707) & (0.040) & (0.395) \\
  & & & & & \\
 Distance & -0.142$\times 10^{-3}$ & 0.470$\times 10^{-4**}$ & 0.154$\times 10^{-5}$ & 0.125$\times 10^{-4}$ & 0.810$\times 10^{-4}$ \\

  & (0.138) & (0.072) & (0.980) & (0.620) & (0.112) \\
  & & & & & \\
 Max. temperature & 0.001 & 0.005 & 0.003 & -0.001 & -0.005 \\
 
  & (0.943) & (0.227) & (0.803) & (0.735) & (0.527) \\
  & & & & & \\
 Min. temperature & 0.009 & -0.006 & -0.007 & 0.005 & 0.007 \\

  & (0.973) & (0.366) & (0.688) & (0.478) & (0.590) \\
  & & & & & \\
 Constant & 0.086 & -0.097 & -0.050 & -0.002 & 0.064 \\

  & (0.729) & (0.160) & (0.764) & (0.972) & (0.627) \\

 District fixed effect & No & No & No & No & No \\
 \hline
 R-squared & 0.260 & 0.255 & 0.038 & 0.524 & 0.321 \\
 \hline
 Number of days detected & 17 & 17 & 17 & 17 & 17 \\
 \hline
 Number of people detected in a status & 899 & 55 & 180 & 62 & 109 \\
 \hline
 Total number of people detected & 1,305 & 1,305 & 1,305 & 1,305 & 1,305 \\
 \hline
 \end{tabular}
\label{table:OLS_I}
\begin{tablenotes}
\small
\item $p$-values in parentheses \newline $^{*}$ $p<0.1$, $^{**}$ $p<0.05$, $^{***}$ $p<0.01$
\end{tablenotes}
\end{threeparttable}
\end{table}
\egroup

\newpage

\bgroup
\def\arraystretch{1.5}
\begin{table}[ht]
\begin{threeparttable}
\centering
\caption{OLS regressions for Phase II}
\begin{tabular}{ C{3.5cm}C{2.1cm}C{2.1cm}C{2.1cm}C{2.3cm}C{2.1cm}  }
 \hline
  & No facial covers & Wearing masks wrongly & Disposable masks & Respirator masks & Other facial covers \\
 \hline
  & (1) & (2) & (3) & (4) & (5) \\
 \hline
 Announcement & -1.641$^{***}$ & 0.215 & 0.811$^{*}$ & 0.353$^{**}$ & 0.262 \\

  & (0.003) & (0.327) & (0.062) & (0.048) & (0.195) \\
  & & & & & \\
 Distance & -0.001$^{***}$ & 0.887$\times 10^{-4}$ & 0.537$\times 10^{-3}$$^{**}$ & 0.308$\times 10^{-3}$$^{***}$ & 0.169$\times 10^{-3}$ \\

  & (0.002) & (0.519) & (0.048) & (0.006) & (0.180) \\
  & & & & & \\
 Announcement $\times$ Distance & 0.001$^{***}$ & -0.867$\times 10^{-4}$ & -0.518$\times 10^{-3}$$^{*}$ & -0.289$\times 10^{-3}$$^{**}$ & -0.176$\times 10^{-3}$ \\

  & (0.002) & (0.535) & (0.061) & (0.011) & (0.170) \\
  & & & & & \\
 Max. temperature & -0.004 & 0.003$^{**}$ & 0.001 & 0.001 & -0.001 \\

  & (0.222) & (0.024) & (0.809) & (0.232) & (0.461) \\
  & & & & & \\
 Min. temperature & 0.007 & -0.007$^{***}$ & 0.001 & -0.002 & 0.001 \\

  & (0.136) & (0.000) & (0.881) & (0.291) & (0.615) \\
  & & & & & \\
 Constant & 1.625$^{***}$ & -0.202 & -0.788$^{*}$ & -0.377$^{**}$ & -0.258 \\

  & (0.004) & (0.362) & (0.072) & (0.036) & (0.204) \\

 District fixed effect & Controlled & Controlled & Controlled & Controlled & Controlled \\
 \hline
 R-squared & 0.302 & 0.195 & 0.206 & 0.297 & 0.100 \\
 \hline
 Number of days detected & 140 & 140 & 140 & 140 & 140 \\
 \hline
 Number of people detected in a status & 12,879 & 1,421 & 5,393 & 1,403 & 1,540 \\
 \hline
 Total number of people detected & 22,636 & 22,636 & 22,636 & 22,636 & 22,636 \\
 \hline
 \end{tabular}
\label{table:OLS_II}
\begin{tablenotes}
\small
\item $p$-values in parentheses \newline $^{*}$ $p<0.1$, $^{**}$ $p<0.05$, $^{***}$ $p<0.01$
\end{tablenotes}
\end{threeparttable}
\end{table}
\egroup

\newpage

\bgroup
\def\arraystretch{1.5}
\begin{table}[ht]
\begin{threeparttable}
\centering
\caption{Marginal effect model across different locations for Phase II}
\begin{tabular}{ C{3.5cm}C{2.1cm}C{2.1cm}C{2.1cm}C{2.3cm}C{2.1cm} }
 \hline
  & No facial covers & Wearing masks wrongly & Disposable masks & Respirator masks & Other facial covers \\
 \hline
  & (1) & (2) & (3) & (4) & (5) \\
 \hline
 \emph{Inside the station} (-220 m) & -1.876$^{***}$ & 0.235 & 0.925$^{*}$ & 0.416$^{**}$ & 0.300 \\

  & (0.003) & (0.348) & (0.061) & (0.040) & (0.190) \\
  & & & & & \\
 \emph{Station entrance} (0 m) & -1.641$^{***}$ & 0.215 & 0.811$^{*}$ & 0.353$^{**}$ & 0.262 \\

  & (0.003) & (0.327) & (0.062) & (0.048) & (0.195) \\
  & & & & & \\
 \emph{Left station} (220 m) & -1.406$^{***}$ & 0.196 & 0.697$^{*}$ & 0.289$^{*}$ & 0.223 \\

  & (0.003) & (0.303) & (0.064) & (0.061) & (0.202) \\
 \hline
 Number of people detected & 12,879 & 1,421 & 5,393 & 1,403 & 1,540 \\
 \hline
 \end{tabular}
\label{table:margeff_II}
\begin{tablenotes}
\small
\item $p$-values in parentheses \newline $^{*}$ $p<0.1$, $^{**}$ $p<0.05$, $^{***}$ $p<0.01$
\end{tablenotes}
\end{threeparttable}
\end{table}
\egroup

\newpage

\bgroup
\def\arraystretch{1.3}
\begin{table}[ht]
\begin{threeparttable}
\centering
\caption{Marginal effects of Announcements 1, 2 and 3 on people's mask-wearing behaviour for Phase II}
\begin{tabular}{ C{3.5cm}C{2.1cm}C{2.1cm}C{2.1cm}C{2.3cm}C{2.1cm}  }
 \hline
  & No facial covers & Wearing masks wrongly & Disposable masks & Respirator masks & Other facial covers \\
 \hline
  & (1) & (2) & (3) & (4) & (5) \\
 \hline
 \textbf{Announcement 1} & & & & & \\

 \emph{Inside the station} (-220 m) & -0.008 & 0.039 & 0.003 & -0.017 & -0.019 \\

  & (0.901) & (0.117) & (0.944) & (0.410) & (0.421) \\
  & & & & & \\
 \emph{Station entrance} (0 m) & -0.014 & 0.034 & 0.011 & -0.013 & -0.019 \\
 
  & (0.796) & (0.117) & (0.798) & (0.463) & (0.356) \\
  & & & & & \\
 \emph{Left station} (220 m) & -0.020 & 0.029 & 0.019 & -0.009 & -0.019 \\

  & (0.675) & (0.137) & (0.629) & (0.562) & (0.304) \\
 
 \textbf{Announcement 2} & & & & & \\

 \emph{Inside the station} (-220 m) & 0.089 & -0.008 & -0.023 & -0.040 & -0.019 \\

  & (0.228) & (0.794) & (0.701) & (0.101) & (0.503) \\
  & & & & & \\
 \emph{Station entrance} (0 m) & 0.083 & -0.005 & -0.024 & -0.038$^{*}$ & -0.016 \\

  & (0.178) & (0.826) & (0.630) & (0.062) & (0.492) \\
  & & & & & \\
 \emph{Left station} (220 m) & 0.077 & -0.003 & -0.025 & -0.036$^{**}$ & -0.013 \\

  & (0.150) & (0.886) & (0.563) & (0.042) & (0.512) \\

\textbf{Announcement 3} & & & & & \\

 \emph{Inside the station} (-220 m) & -0.177 & -0.026 & 0.154$^{*}$ & -0.004 & 0.045 \\

  & (0.108) & (0.564) & (0.081) & (0.915) & (0.282) \\
  & & & & & \\
 \emph{Station entrance} (0 m) & -0.128$^{*}$ & -0.013 & 0.094 & 0.018 & 0.029 \\

  & (0.079) & (0.658) & (0.108) & (0.446) & (0.292) \\
  & & & & & \\
 \emph{Left station} (220 m) & -0.079 & -0.351$\times 10^{-3}$ & 0.034 & 0.032$^{*}$ & 0.013 \\
  & (0.126) & (0.987) & (0.414) & (0.056) & (0.495) \\
 \hline 
 \end{tabular}
\label{table:margeff_123}
\begin{tablenotes}
\small
\item $p$-values in parentheses \newline $^{*}$ $p<0.1$, $^{**}$ $p<0.05$, $^{***}$ $p<0.01$
\end{tablenotes}
\end{threeparttable}
\end{table}
\egroup

\newpage

\bgroup
\def\arraystretch{1.5}
\begin{table}[ht]
\begin{threeparttable}
\centering
\caption{OLS regression for Phase III}
\begin{tabular}{ C{3.5cm}C{2.1cm}C{2.1cm}C{2.1cm}C{2.1cm}C{2.1cm}  }
 \hline
  & No facial covers & Wearing masks wrongly & Disposable masks & Respirator masks & Other facial covers \\
 \hline
  & (1) & (2) & (3) & (4) & (5) \\
 \hline
 Regulation 2 & 0.116 & 0.046 & -0.132$^{*}$ & -0.041 & 0.011 \\

  & (0.268) & (0.341) & (0.095) & (0.351) & (0.807) \\
  & & & & & \\
 Distance & 0.868$\times 10^{-4}$ & -0.127$\times 10^{-3}$ & 0.122$\times 10^{-3}$ & 0.628$\times 10^{-4}$ & -0.144$\times 10^{-3}$ \\

  & (0.753) & (0.323) & (0.550) & (0.593) & (0.236) \\
  & & & & & \\
 Max. temperature & -0.003 & -0.173$\times 10^{-3}$ & 0.005 & -0.311$\times 10^{-4}$ & -0.002 \\

  & (0.781) & (0.973) & (0.538) & (0.995) & (0.713) \\
  & & & & & \\
 Min. temperature & 0.045$^{**}$ & -0.001 & -0.043$^{***}$ & -0.007 & 0.007 \\

  & (0.033) & (0.888) & (0.007) & (0.402) & (0.406) \\
  & & & & & \\
 Constant & -0.321$^{*}$ & 0.029 & 0.260$^{**}$ & 0.042 & -0.009 \\

  & (0.060) & (0.704) & (0.040) & (0.550) & (0.894) \\

 District fixed effect & Controlled & Controlled & Controlled & Controlled & Controlled \\
 \hline
 R-squared & 0.495 & 0.295 & 0.571 & 0.267 & 0.532 \\
 \hline
 Number of days detected & 34 & 34 & 34 & 34 & 34 \\
 \hline
 Number of people detected in a status & 1,674 & 301 & 1,175 & 288 & 316 \\
 \hline
 Total number of people detected & 3,754 & 3,754 & 3,754 & 3,754 & 3,754 \\
 \hline
 \end{tabular}
\label{table:OLS_III}
\begin{tablenotes}
\small
\item $p$-values in parentheses \newline $^{*}$ $p<0.1$, $^{**}$ $p<0.05$, $^{***}$ $p<0.01$
\end{tablenotes}
\end{threeparttable}
\end{table}
\egroup

\newpage

\bgroup
\def\arraystretch{1.5}
\begin{table}[ht]
\centering
\caption{Predictive power comparison between Regression and MLP analyses: Disposable masks}
\begin{tabular}{ |C{2cm}|C{4cm}|C{2cm}|C{2cm}| }
 \hline
 Phase & Comparative statistics & Regression & Neural Network  \\
 \hline \hline
\multirow{5}{*}{I} & Mean squared error (MSE) & 0.000 & \textbf{0.000} \\ \cline{2-4} 
 & Root-mean-square deviation (RMSE) & 0.000 &  \textbf{0.000}   \\ \cline{2-4} 
 & \multicolumn{1}{c|}{Mean absolute error (MAE)} & \multicolumn{1}{c|}{0.972} & \multicolumn{1}{c|}{\textbf{0.152} } \\ \cline{2-4} 
& \begin{tabular}[c]{@{}c@{}}Symmetric Mean\\ Absolute percentage\\ (sMAPE)\end{tabular} & 1.882 & \textbf{1.249} \\ \hline
\multirow{5}{*}{II} & Mean squared error (MSE) & 0.015 & \textbf{0.012} \\ \cline{2-4} 
 & Root-mean-square deviation (RMSE) & 0.121 &  \textbf{0.110}   \\ \cline{2-4} 
 & \multicolumn{1}{c|}{Mean absolute error (MAE)} & \multicolumn{1}{c|}{0.208} & \multicolumn{1}{c|}{\textbf{0.079} } \\ \cline{2-4} 
& \begin{tabular}[c]{@{}c@{}}Symmetric Mean\\ Absolute percentage\\ (sMAPE)\end{tabular} & 1.516 & \textbf{0.415} \\ \hline
\multirow{5}{*}{III} & Mean squared error (MSE) & 0.012 & \textbf{0.008} \\ \cline{2-4} 
 & Root-mean-square deviation (RMSE) & 0.109 &  \textbf{0.091}   \\ \cline{2-4} 
 & \multicolumn{1}{c|}{Mean absolute error (MAE)} & \multicolumn{1}{c|}{0.255} & \multicolumn{1}{c|}{\textbf{0.073} } \\ \cline{2-4} 
& \begin{tabular}[c]{@{}c@{}}Symmetric Mean\\ Absolute percentage\\ (sMAPE)\end{tabular} & 1.292 & \textbf{0.283} \\ \hline
\end{tabular}
\label{table:MLP_vs_Reg}
\end{table}
\egroup

\newpage

\bgroup
\def\arraystretch{1.5}
\begin{table}[ht]
\begin{threeparttable}
\centering
\caption{T-test for the effects of Announcements 1, 2 and 3 on people's mask-wearing behaviour}
\begin{tabular}{ |C{3.5cm}|C{2.1cm}|C{2.1cm}|C{2.1cm}|C{2.1cm}|C{2.1cm}|  }
 \hline
  Null hypothesis & No facial covers & Wearing masks wrongly & Disposable masks & Respirator masks & Other facial covers \\
 \hline
  & (1) & (2) & (3) & (4) & (5) \\
 \hline
 No Announcement = Announcement 1 & 2.709$^{**}$ & -1.864$^{*}$ & 2.465$^{**}$ & 2.557$^{**}$ & -1.036 \\
 \hline
 Announcement 1 = Announcement 2 & 0.362 & -0.834 & -0.413 & 0.421 & 0.135 \\
 \hline
Announcement 2 = Announcement 3 & -2.706$^{**}$ & 2.026$^{*}$ & -2.435$^{**}$ & -1.817$^{*}$ & 0.848 \\
 \hline
 \end{tabular}
\label{table:ttest123}
\begin{tablenotes}
\small
\item Note: The numbers in the table are t-statistics. $p$-values in parentheses. $^{*}$ $p<0.1$, $^{**}$ $p<0.05$, $^{***}$ $p<0.01$
\end{tablenotes}
\end{threeparttable}
\end{table}
\egroup

\newpage

\bgroup
\def\arraystretch{1.2}
\begin{table}[ht]
\centering
\caption{Rank of mask-wearing behaviour}
\begin{tabular}{ |C{6cm}|C{6cm}| }
 \hline
 Behaviour & Rank \\
 \hline
 Not wearing any facial covers & 1 \\
 \hline
 Wearing mask wrongly & 2 \\
 \hline
 Wearing other facial covers & 3 \\
 \hline
 Wearing a disposable mask & 4 \\
 \hline
 Wearing a respirator mask & 5 \\
 \hline
\end{tabular}
\label{table:rank}
\end{table}
\egroup

\clearpage

\bgroup
\def\arraystretch{1.5}
\begin{table}[ht]
\begin{threeparttable}
\centering
\caption{Strictness Index}
\begin{tabular}{ C{3cm}C{3cm}C{3cm}C{3cm}  }
 \hline
  & Phase I & Phase II & Phase III \\
 \hline
  & (1) & (2) & (3) \\
 \hline
 Regulation 1 & 0.006 & & \\

  & (0.901) & & \\
  & & & \\
 Distance & 0.263$\times 10^{-4}$ & -0.188$\times 10^{-3}$ & -0.383$\times 10^{-4}$ \\

  & (0.712) & (0.548) & (0.254) \\
  & & & \\
 Announcement & & -0.026 & \\
 
  & & (0.959) & \\
  & & & \\
 Announcement $\times$ Distance & & 0.407$\times 10^{-5}$ & \\

  & & (0.990) & \\
  & & & \\
 Regulation 2 & & & 0.052$^{***}$ \\

  & & & (0.000) \\
 \hline
 R-squared & 0.576 & 0.345 & 0.804 \\
 \hline
 Number of days detected & 16 & 140 & 34 \\
 \hline
 Total number of people detected & 1,305 & 22,636 & 3,754 \\
 \hline
 \end{tabular}
\label{table:strictness}
\begin{tablenotes}
\small
\item $p$-values in parentheses \newline $^{*}$ $p<0.1$, $^{**}$ $p<0.05$, $^{***}$ $p<0.01$
\end{tablenotes}
\end{threeparttable}
\end{table}
\egroup

\newpage

\bgroup
\def\arraystretch{1.5}
\begin{table}[ht]
\begin{threeparttable}
\centering
\caption{Marginal effect of Strictness Index}
\begin{tabular}{ C{3cm}C{3cm}C{3cm}C{3cm}}
 \hline
  & Announcement 1 & Announcement 2 & Announcement 3 \\
 \hline
  & (1) & (2) & (3) \\
 \hline
 \emph{Inside the station} (-220 m) & -0.372$^{***}$ & -0.128$^{**}$ & 0.223$^{***}$ \\

  & (0.000) & (0.000) & (0.000) \\
  & & & \\
 \emph{Station entrance} 
 
 (0 m) & -0.376$^{***}$ & -0.139$^{***}$ & 0.193$^{***}$ \\

  & (0.000) & (0.000) & (0.000) \\
  & & & \\
 \emph{Left station} 
 
 (220 m) & -0.380$^{***}$ & -0.151$^{***}$ & 0.163$^{***}$ \\

  & (0.000) & (0.000) & (0.000) \\
 \hline
 \end{tabular}
\label{table:marg_strict}
\begin{tablenotes}
\small
\item $p$-values in parentheses \newline $^{*}$ $p<0.1$, $^{**}$ $p<0.05$, $^{***}$ $p<0.01$
\end{tablenotes}
\end{threeparttable}
\end{table}
\egroup

\newpage

\bgroup
\def\arraystretch{1.25}
\begin{table}[ht]
\begin{threeparttable}
\centering
\caption{Marginal effects of Announcements 1, 2 and 3 on people's mask-wearing behaviour for Phase II including district characteristics}
\begin{tabular}{ C{3.5cm}C{2.1cm}C{2.1cm}C{2.1cm}C{2.3cm}C{2.1cm}  }
 \hline
  & No facial covers & Wearing masks wrongly & Disposable masks & Respirator masks & Other facial covers \\
 \hline
  & (1) & (2) & (3) & (4) & (5) \\
 \hline
 \textbf{Announcement 1} & & & & & \\

 \emph{Inside the station} (-220 m) & -0.007 & 0.028 & 0.005 & -0.013 & -0.013 \\

  & (0.908) & (0.235) & (0.907) & (0.492) & (0.525) \\
  & & & & & \\
 \emph{Station entrance} (0 m) & -0.012 & 0.025 & 0.012 & -0.011 & -0.015 \\
 
  & (0.819) & (0.232) & (0.756) & (0.515) & (0.437) \\
  & & & & & \\
 \emph{Left station} (220 m) & -0.017 & 0.022 & 0.019 & -0.009 & -0.016 \\

  & (0.719) & (0.251) & (0.591) & (0.565) & (0.359) \\
 
 \textbf{Announcement 2} & & & & & \\

 \emph{Inside the station} (-220 m) & 0.071 & 0.003 & -0.027 & -0.034 & -0.013 \\

  & (0.305) & (0.925) & (0.616) & (0.136) & (0.605) \\
  & & & & & \\
 \emph{Station entrance} (0 m) & 0.081 & -0.001 & -0.032 & -0.036$^{*}$ & -0.011 \\

  & (0.172) & (0.953) & (0.481) & (0.065) & (0.600) \\
  & & & & & \\
 \emph{Left station} (220 m) & 0.090$^{*}$ & -0.005 & -0.038 & -0.038$^{**}$ & -0.009 \\

  & (0.085) & (0.796) & (0.353) & (0.029) & (0.619) \\

\textbf{Announcement 3} & & & & & \\

 \emph{Inside the station} (-220 m) & -0.233$^{**}$ & -0.011 & 0.171$^{**}$ & 0.019 & 0.053 \\

  & (0.034) & (0.803) & (0.041) & (0.597) & (0.188) \\
  & & & & & \\
 \emph{Station entrance} (0 m) & -0.171$^{**}$ & -0.002 & 0.113$^{**}$ & 0.028 & 0.032 \\

  & (0.019) & (0.947) & (0.045) & (0.236) & (0.226) \\
  & & & & & \\
 \emph{Left station} (220 m) & -0.109$^{**}$ & 0.007 & 0.053 & 0.038$^{**}$ & 0.012 \\
  & (0.033) & (0.728) & (0.180) & (0.025) & (0.530) \\
 \hline 
 \end{tabular}
\label{table:margin_district}
\begin{tablenotes}
\small
\item $p$-values in parentheses \newline $^{*}$ $p<0.1$, $^{**}$ $p<0.05$, $^{***}$ $p<0.01$
\end{tablenotes}
\end{threeparttable}
\end{table}
\egroup

\newpage

\bgroup
\def\arraystretch{1.5}
\begin{table}[ht]
\begin{threeparttable}
\centering
\caption{Robust test for Regulation 2 and Announcement effects on mask-wearing behaviour}
\begin{tabular}{ C{3.5cm}C{2.1cm}C{2.1cm}C{2.4cm}C{2.4cm}C{2.1cm}  }
 \hline
  & No facial covers & Wearing masks wrongly & Disposable masks & Respirator masks & Other facial covers \\
 \hline
  & (1) & (2) & (3) & (4) & (5) \\
 \hline
Announcement & -1.844$^{***}$ & 0.273 & 0.874$^{**}$ & 0.430$^{**}$ & 0.266 \\

  & (0.001) & (0.207) & (0.038) & (0.017) & (0.167) \\
  & & & & & \\
 Announcement $\times$ Distance & 0.001$^{***}$ & -0.132$\times 10^{-3}$ & -0.572$\times 10^{-3}$$^{**}$ & -0.350$\times 10^{-3}$$^{***}$ & -0.189$\times 10^{-3}$ \\

  & (0.000) & (0.330) & (0.031) & (0.002) & (0.119) \\
  & & & & & \\
 Regulation 2 & 0.135 & 0.024 & -0.142$^{*}$ & -0.053 & 0.036 \\

  & (0.172) & (0.614) & (0.058) & (0.201) & (0.448) \\
  \hline
 District characteristics & Controlled & Controlled & Controlled & Controlled & Controlled \\
 \hline
 \end{tabular}
\label{table:Robust_district}
\begin{tablenotes}
\small
\item $p$-values in parentheses \newline $^{*}$ $p<0.1$, $^{**}$ $p<0.05$, $^{***}$ $p<0.01$
\end{tablenotes}
\end{threeparttable}
\end{table}
\egroup

\newpage

\bgroup
\def\arraystretch{1.5}
\begin{table}[ht]
\begin{threeparttable}
\centering
\caption{Robust test for seasonal effects}
\begin{tabular}{ C{3.5cm}C{2.1cm}C{2.1cm}C{2.4cm}C{2.4cm}C{2.1cm}  }
 \hline
  & No facial covers & Wearing masks wrongly & Disposable masks & Respirator masks & Other facial covers \\
 \hline
  & (1) & (2) & (3) & (4) & (5) \\
 \hline
 Regulation 1 & -0.012 & -0.076 & 0.053 & 0.130$^{**}$ & -0.095 \\

 & (0.956)& (0.190) & (0.707) & (0.040) & (0.395) \\
& & & & & \\
Announcement & -1.599$^{***}$ & 0.176 & 0.778$^{*}$ & 0.355$^{**}$ & 0.290 \\

  & (0.003) & (0.414) & (0.072) & (0.046) & (0.151) \\
  & & & & & \\
 Announcement $\times$ Distance & 0.001$^{***}$ & -0.680$\times 10^{-4}$ & -0.504$\times 10^{-3}$$^{*}$ & -0.292$\times 10^{-3}$$^{**}$ & -0.191$\times 10^{-3}$ \\

  & (0.002) & (0.619) & (0.066) & (0.010) & (0.138) \\
  & & & & & \\
 Regulation 2 & 0.116 & 0.046 & -0.132$^{*}$ & -0.041 & 0.011 \\

  & (0.268) & (0.341) & (0.095) & (0.351) & (0.807) \\
  \hline
 Control variables & Controlled & Controlled & Controlled & Controlled & Controlled \\
 \hline
 Seasonal fixed effect & Controlled & Controlled & Controlled & Controlled & Controlled \\
 \hline
 \end{tabular}
\label{table:Robust_season_overall}
\begin{tablenotes}
\small
\item $p$-values in parentheses \newline $^{*}$ $p<0.1$, $^{**}$ $p<0.05$, $^{***}$ $p<0.01$
\end{tablenotes}
\end{threeparttable}
\end{table}
\egroup

\newpage

\bgroup
\def\arraystretch{1.4}
\begin{table}[ht]
\centering
\caption{Predictive power comparison of multilayer perceptrons across different memory windows: disposable masks}
\begin{tabular}{ |C{2cm}|C{4cm}|C{2cm}|C{2cm}| C{2cm}| C{2cm}| }
 \hline
 Phase & Comparative statistics & 0 days & 3 days & 7 days & 11 days  \\
 \hline \hline
\multirow{5}{*}{I} & Mean squared error (MSE) & \textbf{0.000} & 0.000   & 0.000  & 0.000  \\ \cline{2-6} 
                   & Root-mean-square deviation (RMSE)  & \textbf{0.000}  & 0.000   & 0.000 & 0.000 \\ \cline{2-6} 
                   & Mean absolute error (MAE) & \textbf{0.152} & 0.207  & 0.200  & 0.197  \\ \cline{2-6} 
                   & Symmetric Mean Absolute Percentage Error (sMAPE)  & \textbf{1.249} & 1.363  & 1.299 & 1.382  \\ \hline
\multirow{5}{*}{II} & Mean squared error (MSE) & \textbf{0.012} & 0.014   & 0.015  & 0.014  \\ \cline{2-6} 
                   & Root-mean-square deviation (RMSE)  & \textbf{0.110}  & 0.118   & 0.123 & 0.119 \\ \cline{2-6} 
                   & Mean absolute error (MAE) & \textbf{0.079} & 0.087  & 0.090  & 0.090  \\ \cline{2-6}  
                   & Symmetric Mean Absolute Percentage Error (sMAPE)  & \textbf{0.415} & 0.435  & 0.439 & 0.437  \\ \hline
\multirow{5}{*}{III} & Mean squared error (MSE) & \textbf{0.008} & 0.009   & 0.010  & 0.010  \\ \cline{2-6} 
                   & Root-mean-square deviation (RMSE)  & \textbf{0.091}  & 0.097   & 0.100 & 0.101 \\ \cline{2-6} 
                   & Mean absolute error (MAE) & \textbf{0.073} & 0.076  & 0.078  & 0.077  \\ \cline{2-6}  
                   & Symmetric Mean Absolute Percentage Error (sMAPE)  & \textbf{0.283} & 0.288  & 0.297 & 0.286  \\ \hline
\multirow{5}{*}{Overall} & Mean squared error (MSE) & \textbf{0.010} & 0.012   & 0.012  & 0.012  \\ \cline{2-6} 
                   & Root-mean-square deviation (RMSE)  & \textbf{0.101}  & 0.108   & 0.111 & 0.111 \\ \cline{2-6} 
                   & Mean absolute error (MAE) & \textbf{0.074} & 0.080  & 0.083  & 0.085  \\ \cline{2-6} 
                   & Symmetric Mean Absolute Percentage Error (sMAPE)  & \textbf{0.540} & 0.556  & 0.555 & 0.557  \\ \hline
\end{tabular}
\label{table:NN_Memory window}
\end{table}
\egroup

\newpage

\bgroup
\def\arraystretch{1.3}
\begin{table}[ht]
\centering
\caption{Predictive power comparison between multilayer perceptrons and random forest: disposable masks }
\begin{tabular}{ |C{2cm}|C{4cm}|C{2cm}|C{2cm}| }
 \hline
 Phase & Comparative statistics & Multilayer Perceptrons & Random Forest  \\
 \hline \hline
\multirow{5}{*}{I} & Mean squared error (MSE) & \textbf{0.000} & 0.000 \\ \cline{2-4} 
 & Root-mean-square deviation (RMSE) & \textbf{0.000} & 0.000  \\ \cline{2-4} 
 & \multicolumn{1}{c|}{Mean absolute error (MAE)}  & \multicolumn{1}{c|}{\textbf{0.152} } & \multicolumn{1}{c|}{0.181}\\ \cline{2-4} 
& \begin{tabular}[c]{@{}c@{}}Symmetric Mean\\ Absolute percentage\\ (sMAPE)\end{tabular}  & \textbf{1.249} & 1.340 \\ \hline
\multirow{5}{*}{II} & Mean squared error (MSE)  & \textbf{0.012} & 0.016\\ \cline{2-4} 
 & Root-mean-square deviation (RMSE)  &  \textbf{0.110} & 0.127   \\ \cline{2-4}
 & \multicolumn{1}{c|}{Mean absolute error (MAE)}  & \multicolumn{1}{c|}{\textbf{0.079} } & \multicolumn{1}{c|}{0.093} \\ \cline{2-4} 
& \begin{tabular}[c]{@{}c@{}}Symmetric Mean\\ Absolute percentage\\ (sMAPE)\end{tabular}  & \textbf{0.415} & 0.456 \\ \hline
\multirow{5}{*}{III} & Mean squared error (MSE) & \textbf{0.008} & 0.011 \\ \cline{2-4} 
 & Root-mean-square deviation (RMSE)  &  \textbf{0.091} & 0.106   \\ \cline{2-4} 
 & \multicolumn{1}{c|}{Mean absolute error (MAE)}  & \multicolumn{1}{c|}{\textbf{0.073} } & \multicolumn{1}{c|}{0.105} \\ \cline{2-4} 
& \begin{tabular}[c]{@{}c@{}}Symmetric Mean\\ Absolute percentage\\ (sMAPE)\end{tabular}  & \textbf{0.283} & 0.403 \\ \hline
\multirow{5}{*}{Overall} & Mean squared error (MSE)  & \textbf{0.010} & 0.015\\ \cline{2-4} 
 & Root-mean-square deviation (RMSE)  &  \textbf{0.101} & 0.121   \\ \cline{2-4}
 & \multicolumn{1}{c|}{Mean absolute error (MAE)}  & \multicolumn{1}{c|}{\textbf{0.074} } & \multicolumn{1}{c|}{0.091} \\ \cline{2-4} 
& \begin{tabular}[c]{@{}c@{}}Symmetric Mean\\ Absolute percentage\\ (sMAPE)\end{tabular}  & \textbf{0.540} & 0.592 \\ \hline
\end{tabular}
\label{table:NN_VS_RF}
\end{table}
\egroup

\newpage

\bgroup
\def\arraystretch{1.5}
\begin{table}[ht]
\centering
\caption{Random forest vs. multilayer perceptrons: time consumption and fitting nodes}
\begin{tabular}{ |C{4cm}|C{2.4cm}|C{2 cm}|}
 \hline
  & RF & MLP  \\
 \hline \hline
 Time consumption & 0.94 s & 0.84 s \\
 \hline
 Fitting nodes & 49,995 & 50 \\
 \hline
\end{tabular}
\label{table:RF_vs_MLP_time_node}
\end{table}
\egroup

\clearpage

\clearpage

\clearpage

\clearpage

\end{document}